# The Settling of Warped Disks in Oblate Dark Halos

JOHN DUBINSKI[1] AND KONRAD KUIJKEN[2,3]

Harvard–Smithsonian Center for Astrophysics, 60 Garden St., Cambridge, MA 02138

April 5, 1994

## ABSTRACT

When a galaxy forms, the disk may initially be tilted with respect to a flattened dark halo. The misalignment between the disk and the halo is a common explanation for galactic disk warps, since in this state, disks have precessing bending modes which resemble real warps. A dubious assumption in this theory is that the gravitational response of the halo is negligible. We therefore calculate the response of an oblate halo to a precessing inclined exponential disk using a variety of techniques. We construct models with a rigid exponential disk precessing in a particle halo, a particle disk precessing inside a static bulge/halo potential, and a self-consistent model with a particle disk, bulge and halo. When the disk:halo mass ratio is small ($\sim 10\%$) within 5 exponential scale radii, the disk settles to the equatorial plane of the halo within 5 orbital times. When the disk and halo mass are comparable, the halo rapidly aligns with the disk within a few orbital times, while the disk inclination drops. The rapid response of the halo to a inclined disk suggests that the warps seen in galactic disks are not likely the bending modes of a precessing disk inside of triaxial halos. If a galaxy forms inclined to the principal plane of a dark halo, either the disk will settle to a principal plane or the inner halo will twist to align with the disk. The outer halo will remain misaligned for a much longer time and therefore may still exert a torque. Warped bending modes may still exist if the misalignment of the outer halo persists for a Hubble time.

*Subject headings:* galaxies: kinematics and dynamics – galaxies: structure – galaxies: spiral – galaxies: formation — methods: numerical

## 1 INTRODUCTION

Galactic disks observed edge-on are often warped, resembling cosmic integral signs or old gramophone records left on the radiator. While such warps are a common phenomenon in the local universe with a seemingly simple geometry, they have evaded a satisfying physical explanation. The frequency of warped disks is very large: at least half of galaxies show large-scale bending, both in HI (Bosma 1991) and at optical wavelengths (Sánchez-Saavedra et al. 1990). Perhaps all galaxies are warped at some level if looked at closely enough. There are various physical explanations for such warps: for reviews, see Toomre (1983), Casertano (1990), and Binney (1992). In this paper we will not consider the possibility that warps are a magnetic phenomenon, since it appears that the required cosmic magnetic field strengths are too high (Binney 1991). However, we note that there remains the curious effect that galaxy warps display considerable large-scale alignment (Battaner et al. 1991), which is difficult to explain with other mechanisms.

---

[1]I: dubinski@cfa.harvard.edu

[2]Hubble Fellow

[3]I: kuijken@cfa.harvard.edu

The most compelling steady-state theories for warps postulate the existence of a tidal field on the disk caused by the misalignment between the disk and the symmetry planes of a presumed, flattened dark halo. A tilted disk in a flattened halo feels a gravitational torque, and since the disk has intrinsic spin, it responds to the torque by precessing like a top. The precession frequency in the halo is a function of radius, and so, in the absence of self-gravity a warped disk would wind up over time.[1] When the effects of self-gravity are taken into account, though, realistic warped configurations exist in which the galaxy precesses slowly like a solid body inside the halo. These configurations can be viewed as a neutral mode of vertical oscillation of the disk in the background halo potential (Toomre 1983; Dekel & Shlosman 1983; Sparke 1984; Sparke & Casertano 1988; Kuijken 1991). Even if a disk does not form exactly in a warped mode within a halo it will eventually settle to the mode within a Hubble time (Hofner & Sparke 1993). The precession of the warp mode is retrograde if the halo is oblate, with frequency of the order of $\epsilon V_c/h$, where $\epsilon$ is the halo ellipticity, $V_c$ is the circular speed and $h$ is the disk exponential scale length.

Ostriker & Binney (1989) have also proposed the steady accretion of matter from cosmological infall as an explanation for warps: the accreted material can add angular momentum to the disk or halo, causing the disk to slew slowly inside the halo and develop a warp.

Recent work in the theory of galaxy formation supports aspects of both of these models. In models such as the cold dark matter (CDM) cosmology, which invoke collisionless dark matter as a major constituent of the universe, galaxy formation occurs during the simultaneous collapse of the gas and the dark matter. During the collapse of the seed perturbation, gas dissipates and settles into a disk inside a halo of dark matter (White & Rees 1978), eventually turning into a spiral galaxy. The dark halos in cosmological models containing collisionless dark matter are triaxial and often highly flattened (Frenk et al 1988; Dubinski & Carlberg 1991; Warren et al. 1992). When the disk finally cools and settles within the halo, the disk angular momentum may not necessarily be aligned with the principal axes of the halo, and so the disk will often be inclined with respect to the halo principal planes. This misalignment of the disk and dark halo angular momenta is seen in simulations of collapsing gas and collisionless dark matter (Katz & Gunn 1991). The disks would then soon find their warp modes. In some sense, warps may be fossils of the messy collapse inherent

---

[1] In the outer regions of a disk the winding timescale can be quite long (Tubbs & Saunders 1979), though, and if the halo flattening has the appropriate variation with radius no winding may occur at all (Petrou 1980).

to hierarchical galaxy formation.

Continued accretion of matter ('secondary infall') onto disk galaxies is also expected in many (especially high-$\Omega$) cosmologies (e.g. Gunn 1977; Ryden & Gunn 1987; Frenk et al. 1988). Late accretion is usually in the form of large lumps of material instead of a uniform rain (e.g. Kauffmann & White 1993). Tidal interactions of nearby density peaks naturally produce a random component in the angular momentum of infalling material, and it is therefore expected that infalling material is not aligned with the disk plane of a galaxy (e.g. Quinn & Binney 1992). Consequently, infall must lead to angular momentum realignment in the disk or halo, and hence to transient warping. If secondary infall with time-varying angular momentum is continuous, so is warping of disk galaxies. There is plenty of observational evidence that infall of material with misaligned angular momentum occurs in reality: from polar rings (e.g. Whitmore, McElroy, & Schweizer 1987; Sackett & Sparke 1991) counter-rotating cores in ellipticals (e.g. Franx & Illingworth 1988; Bender 1988; Jedrzejewski & Schechter 1988) and from misaligned or counter-rotating gas and stellar disks in spiral galaxies (e.g. Bettoni, Fasano & Galletta 1990, Rix et al 1992, Bertola et al. 1992, Braun, Walterbos & Kennicut 1992, Merrifield & Kuijken 1994).

The bending mode theory of warps overlooks some potentially important physical effects which have been alluded to but not explored in detail. Toomre (1983) pointed out that a precessing disk inside an oblate halo will feel dynamical friction and therefore will eventually settle to the equatorial plane. Dekel & Shlosman (1983) suggested that the time scale may be as long as a Hubble time and therefore are not important, though there have been no further calculations to support this. Binney (1992) also discussed how gravitational wakes induced by the disk will resist its precession. Recent work on the warp modes (Sparke & Casertano 1988; Kuijken 1991; Hofner & Sparke 1993) are all based on the assumption of a static halo potential. This assumption has never been properly justified. Dubinski (1994) has recently shown that the shape of the dark halo changes significantly in response to a disk forming in a principal plane (see also Flores et al 1993). The shape changes as the orbits of the halo particles adiabatically become aware of the rounder potential. Clearly, in the more general case of a tilted, precessing disk inside a flattened halo a similar modification of the halo will occur.

Our goal is to examine the response of the flattened halo to an embedded tilted, precessing, warped disk, and hence examine the implications for the lifetime of such a warped 'mode'. We tackle this problem in different ways. In §2, we construct models including a rigid-body exponential disk and a dynamic, flattened N-body halo. The disk is tilted inside the halo and set spinning at a rate expected for real galaxies. The equations of motion for the precessing rigid disk and the trajectories of the particles in the halo are solved simultaneously. We find that the disk and halo rapidly align. In §3, we describe methods for setting up an equilibrium disk/bulge/halo N-body system. We then simulate a tilted N-body disk inside a static flattened bulge/halo potential and demonstrate the stability and existence of warped modes as calculated by Sparke & Casertano (1988). We then make the bulge and halo dynamic and discover that as was the case with a solid disk in a dynamic halo, the disk, bulge and halo align within a few orbital periods. In §4, we examine the problems these results raise for theories using warped modes and discuss alternatives. In §5, we summarize our results and their implications.

## 2  Rigid Exponential Disk

The main function of self-gravity in a warped-mode disk is to preserve the disk against differential precession in the oblate halo potential. In our first experiments, we make self-gravity infinitely strong, and treat the disk as a single, flattened, solid body. We give it an exponential density profile, so its gravitational effect on the halo is similar to that of a real stellar disk. Since warped disks are still quite flat, this setup allows us to examine the response of the halo to an embedded precessing disk while avoiding the difficulties of setting up and simulating a stable, collisionless disk. We will tackle the full problem in the next section. The dynamics of the solid disk are treated by solving Euler's equations of motion for a rigid body (Goldstein 1980) simultaneously with the dynamics of the N-body halo. The halo particles therefore move in the combined potential of the precessing rigid disk and their own gravity while the disk precesses under the influence of the torque from the particle halo.

We take the disk to be a homoeoid (Binney and Tremaine 1987) with flattening $q = 0.07$ in the density with an exponential profile in the elliptical radial coordinate truncated at 4.5 radial scale radii. In practice, we calculated the components of the force and potential on a $R - z$ grid and interpolated for values between the grid points.

For the halo potential we use a lowered Evans (1993) model (Kuijken & Dubinski 1994) with the parameters shown in Table 1 (also see appendix). These self-consistent models have a finite radius and an analytic two-integral distribution function, making them simple to realize as $N$-body systems. The halo is realized with 50,000 particles. We do not include a bulge for the rigid disk models. Figure 1 shows the rotation curve of these disk/halo models. The contribution of the disk to the total mass of the system is small in comparison to realistic galaxies but the model provides a low-mass control of some stronger effects which we will discuss below. The spin frequency of the rigid disk is determined by calculating the total angular momentum in the disk according to the rotation curve of the dark halo and dividing by the $I_{zz}$ component of the moment of inertia. The resulting precession frequency for these disks is,

$$\omega_p = -\frac{1}{I_{zz}\omega_z \sin\theta}\frac{\partial V}{\partial \theta} \qquad (1)$$

where $V$ is the potential, $-\partial V/\partial \theta$ is the torque, $\theta$ is the inclination angle and $\omega_z$ is the spin frequency (Goldstein 1980). For the model discussed here, $\omega_z = 0.40$, and $\omega_p = -0.050$. The precession frequency is typical of a fast warp mode in flattened halos (see below). We calculate the torque on the exponential disk by summing over the torque on the particles by the disk. Since there are no external torques, by Newton's 3rd law the net torque on the disk is just the negative of the torque on the particles. The N-body forces are calculated with a tree code (Dubinski 1988; Barnes & Hut 1986) using an opening angle criterion $\theta = 1.0$ and quadrupole order forces. We integrated the equations with a second order predictor-corrector because of the velocity terms in the rigid body equations of motion.

As a test of the program, we set up a massless disk inside the particle halo. As ex-pected, the disk precessed at the constant rate predicted by eq. 1 and maintained a constant inclination.

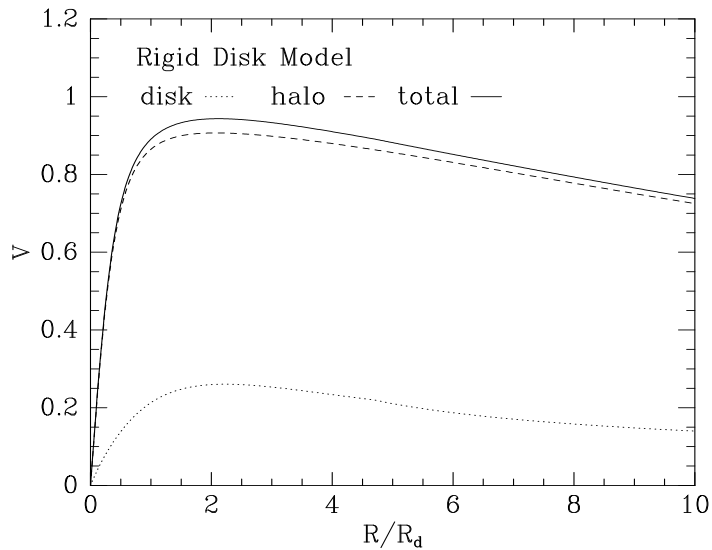

Fig. 1.—The rotation curve in the halo+rigid disk model. The solid line corresponds to the total potential, and the dashed (dotted) lines indicate the contribution from the halo (disk). Compared with conventional disk-halo decompositions for large disk galaxies, the disk mass is unrealistically low.

We then simulated a massive disk's motion in a live halo, in a sequence of three runs with different initial inclinations: $i = 30°$, $45°$, and $60°$. Figures 2 and 3 show the time evolution of the components of the inclination and the longitude of the line of nodes for the three models. In each case, the disk settles rapidly towards the equatorial plane of the halo within 5 spin periods. The final inclination of the disk is not quite zero, since the inner halo twists upwards to meet the disk during the settling phase. The disk wobbles as it interacts with the halo, which is itself adjusting to the presence of the disk. The changes in the precession rate result from the change in inclination as the disk wobbles.

The initial setup is clearly out of equilibrium and the disk and halo relax to a state in which they are aligned. We can understand the process driving the evolution in these simulations as a form of dynamical friction (though the geometry is more complicated than the usual picture of a simple point mass moving through a sea of particles), as follows. As the disk precesses within the halo, it creates a gravitational wake. The wake exerts a torque on the disk that resists the precession. The internal spin of the disk is not changed, so the net effect is to cause the disk to settle into the equatorial plane. Thus, energy is transferred from the disk precession to the surrounding halo particles. Figure 4 shows the twisting of the inner halo three spin periods after the beginning of the $i = 60°$ simulation ($t = 50$). The halo orbits within five scale radii are clearly affected by the precessing disk and are twisting

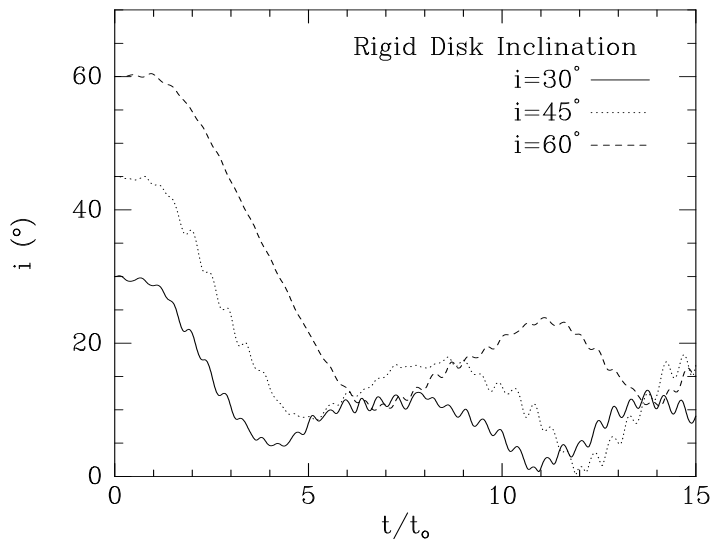

Fig. 2.—The evolution of the inclination of the rigid exponential disk models. The disk settles to the equatorial plane within 4-6 orbital periods and then starts to wobble as it interacts with the gravitational wake created inside the halo. The small scale oscillations are fast nutation of the rigid disk which would not occur in real stellar disks.

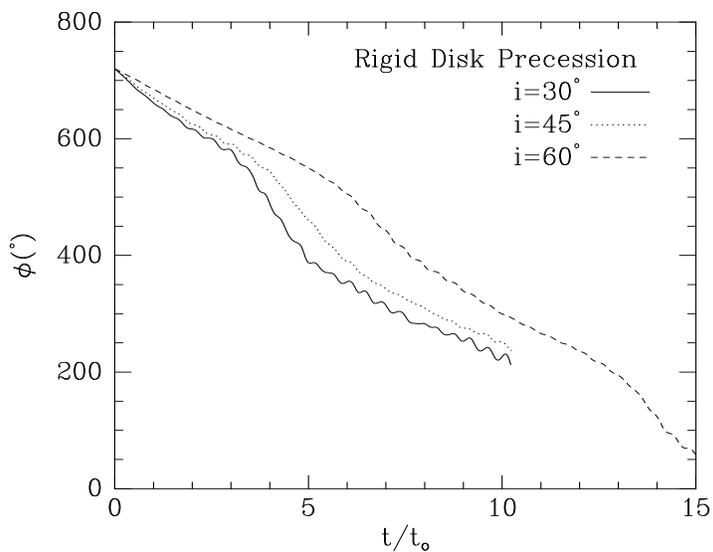

Fig. 3.—Precession of the rigid disk models as shown by the evolution of the disk line of nodes, $\phi$. The precession rate varies with $\cos i$ as the disk wobbles.

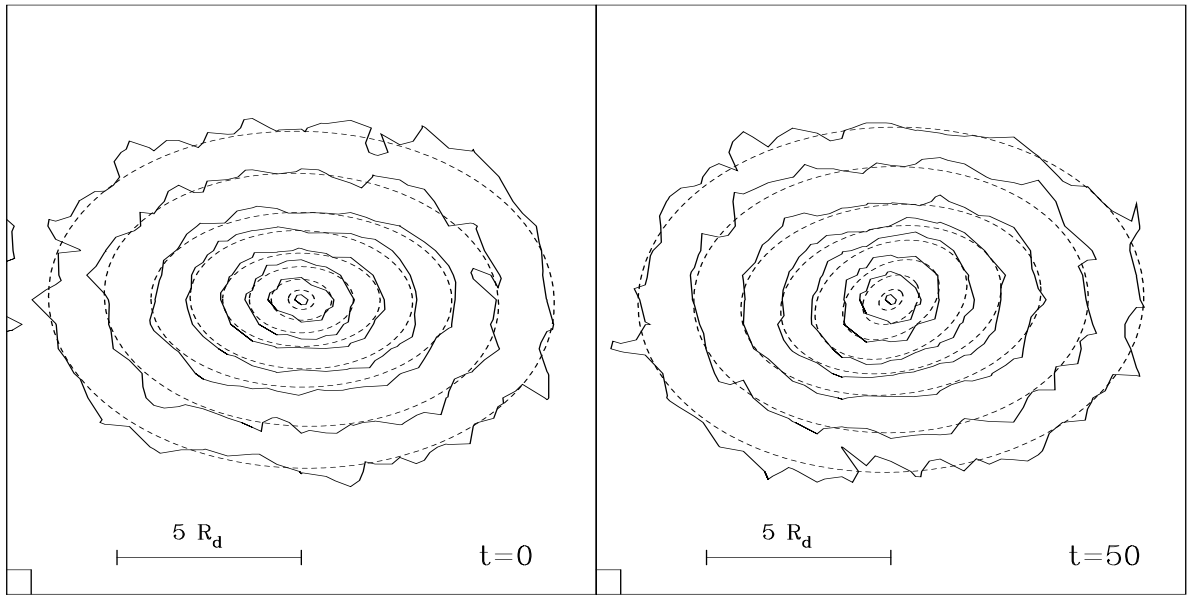

FIG. 4.–The gravitational wake induced by a precessing disk as revealed by the twisting of the inner halo. The measured surface density contours (solid) and elliptical fits to them (dotted) are shown for the model at $t = 0$ and $t = 50$ ($\approx 3$ orbital periods or half a precession period). Within 5 scale radii the inner halo twists up to meet the disk, while the radial density profile is essentially unchanged.

upward to meet the falling disk.

A dynamical friction argument such as this applies to low mass objects which only slightly perturb the halo. While this is true for the disk simulations shown above, real galaxy disks probably have larger disk to halo mass ratios, and a more careful analysis is required to derive an accurate settling time. Despite the unrealistically small disk mass used in the simulations, the time scale for the disk to settle within the halo is only $\approx 5$ dynamical times showing that inclined disks within halos will raise serious wakes, and will quickly try to relax to a state in which they are aligned. In the next section, we examine more realistic galaxy models with more massive disks and show that indeed this is the case.

## 3 PARTICLE DISKS

The modelling of the disk as a rigid body is not necessarily valid, especially in the case of low mass disks where the self-gravity is not sufficient to hold the disk together. Also, the equilibrium shape of the disk is warped and not flat as in the rigid disk simulations. We therefore set up a fully self-consistent system with a particle disk. Our strategy was first to set up an equilibrium system consisting of a disk, a bulge, and a flattened halo placed in the equatorial plane and then, as in the rigid disk models, to tilt the disk. The bulge was included to make the system more realistic and also to suppress the formation of a bar in the N-body disk.

### 3.1 Setting Up an Equilibrium Disk, Bulge and Halo

In the last section, we constructed a self-consistent model of a flattened dark halo only and added a low mass disk. We now describe a method for including the disk and bulge self-consistently, which is an extension of the lowered Evans models (Kuijken & Dubinski 1994—henceforth KD).

In KD, we specified an analytic halo distribution function, $f_h(E, L_z)$, (in this case assumed axisymmetric). The space density $\rho_h(R, z)$ was found by integrating the distribution function over the appropriate portion of phase space,

$$\rho_h(R,z) = \int f_h \, d^3v = \frac{2\pi}{R} \int_{E \geq \Psi + \frac{1}{2}(L_z/R)^2} f_h(E, L_z) \, dL_z \, dE \equiv \rho_h(\Psi, R), \qquad (2)$$

and then solving for the potential, $\Psi$, using Poisson's equation,

$$\nabla^2 \Psi = 4\pi G \rho_h(\Psi, R). \qquad (3)$$

Ideally, we could modify the procedure by also specifying distribution functions for the disk and the bulge, $f_d(E, L_z)$ and $f_b(E, L_z)$, say, with similarly defined densities $\rho_d(\Psi, R)$ and $\rho_b(\Psi, R)$. The net potential, to which all three components now contribute, is then found by solving

$$\nabla^2 \Psi = 4\pi G[\rho_d(\Psi, R) + \rho_b(\Psi, R) + \rho_h(\Psi, R)]. \qquad (4)$$

The difficulty is finding distribution functions which give plausible disk, bulge and halo density distributions. We found in practice that the lowered Evans model distribution function and a King model with a lower energy cutoff were adequate distribution functions for the halo and bulge and produced plausible density profiles for the combined models. Making a disk is more problematic since the vertical and radial velocity dispersions $\sigma_R$ and $\sigma_z$ in two-integral models are per force equal, unlike in realistic disks where typically $\sigma_R = 2\sigma_z$. The vertical velocity dispersion is determined by the disk surface mass density and thickness, so a two-integral disk will have a radial velocity dispersion, and a Toomre $Q$, a factor of $\sim 2$ below realistic values. Two-integral disks with realistic mass and thickness are therefore grossly unstable.

We contented ourselves with a more approximate, but more realistic, disk distribution function. We use the disk density

$$\rho_d(R,z) = \frac{M_d}{4\pi R_d^2 z_d} \exp(-R/R_d) \text{sech}(z/z_d)^2, \qquad (5)$$

truncated at five exponential scale lengths. The solution of equation (4) gives the net potential of the system. In practice, we solve this equation using the iterative technique involving the spherical harmonic expansion of the potential described in KD. Even where the distribution function for the halo and bulge is isotropic, these components are still flattened because of the disk potential.

It now remains to realize the model as a collection of $N$-body particles. For the halo and bulge, which have analytic distribution functions, this is accomplished with the method described by KD. For the disk, we follow the procedure of Hernquist (1993; see Appendix). The Gaussian velocity distributions generated by this recipe, though not dynamically consistent, quickly phase-mix to equilibrium.

All three components are initially in equilibrium when the disk is placed in the equatorial plane of the halo. In an $N$-body simulation of a disk in the equatorial plane we found the density profiles of the disk, bulge and halo did not change significantly over several orbital periods. The only significant evolution consisted in the usual spiral instabilities in the disk from swing amplification of the particle discreteness noise, and a bar mode which began to form in models with a more massive disk. We were therefore confident that the three-component systems were a good initial equilibrium.

For some of the experiments, we also calculated the linear warping mode (Sparke & Casertano 1988; Kuijken 1991) of the system for a moderate inclination ($i \approx 30°$ at the extremity) and tilted the particle disk accordingly. The three components are slightly out of equilibrium initially (since the halo and bulge densities were calculated assuming that the disk lies in the equatorial plane) but this might be expected in a real system; and real systems out of equilibrium are expected to settle into the mode shape (Hofner & Sparke 1993). The mode calculation also gives the precession frequency of the warped disk.

We examined three systems with different halo flattenings and disk:halo mass ratios. Model A has a moderately flattened halo with a disk:halo mass ratio of 1:1 within 5 scale radii. Model B has a nearly spherical but slightly flattened halo with a disk:halo mass ratio of 2:1. Model C has the same flattening as Model A with the disk:halo mass ratio of Model B. Table I summarizes the model parameters. The parameters were chosen to represent cases of rapidly and slowly precessing disks, and different disk masses representative of real galaxies. For convenience, we rescaled the simulations for the models so that the exponential scale radius $R_d$ was unity, and the rotation curve levels off at $v \sim 1.0$ (we multiplied the masses by 10). With this rescaling, the orbital period at the disk half mass radius ($R = 1.7R_d$) is $T_o = 13.4$ in simulation units. Henceforth, we refer to this as the disk orbital period or dynamical time. The precession periods calculated for the warping modes are $T_{p,A} = 129$, $T_{p,B} = 413$, and $T_{p,C} = 143$ corresponding to between 10 and 31 orbital periods. Figures 5 and 6 show the contribution to the density profiles and resulting rotation curves of the three components in Models A and B (C resembles B). In the spirit of a rotation curve conspiracy theory, we purposely choose the three components in suitable ratios to give nearly flat rotation curves.

We represent the disk, bulge and halo with 40,000, 10,000, and 50,000 particles respectively for a total of 100,000 particles. The simulations are run with an N-body tree code using an opening angle criterion, $\theta = 0.9$ with forces calculated to quadrupole order. The orbits are integrated with a leap frog method with a constant timestep, $\Delta t = 0.1$ until $t = 120$, corresponding to 9 orbits at the disk half mass radius.

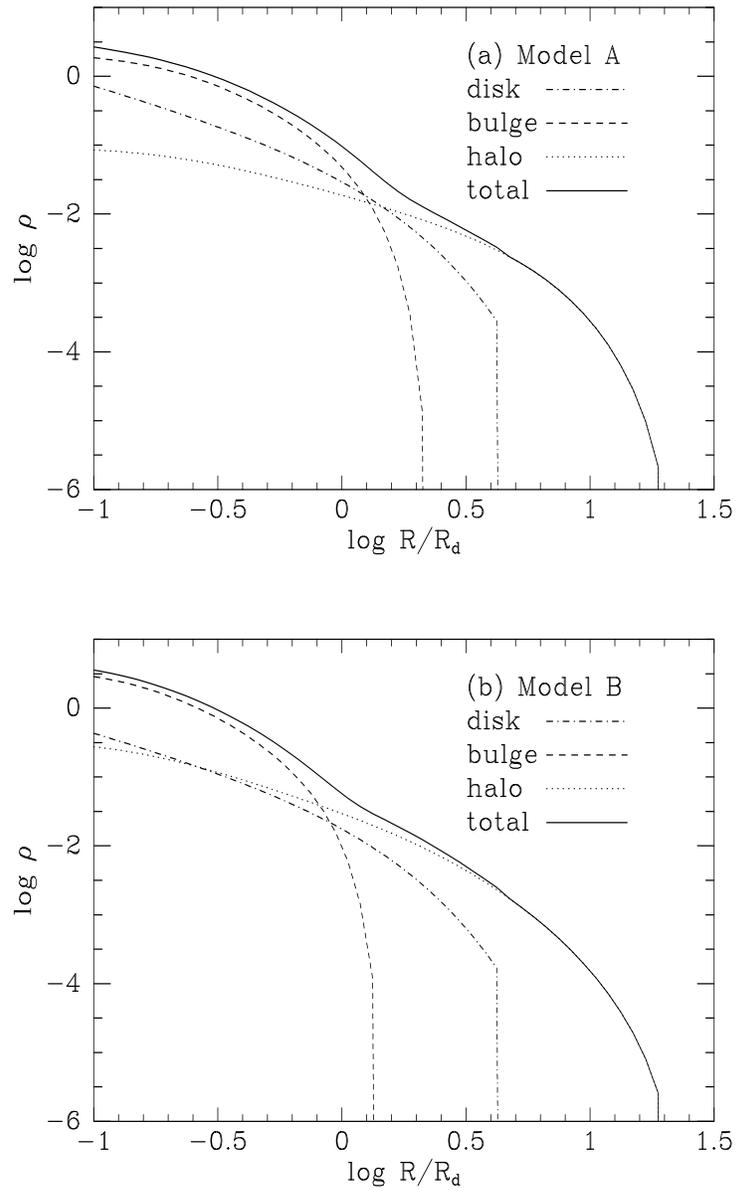

FIG. 5.–Density profiles averaged over spherical shells of the disk, bulge and halo components of (a) Model A and (b) Model B

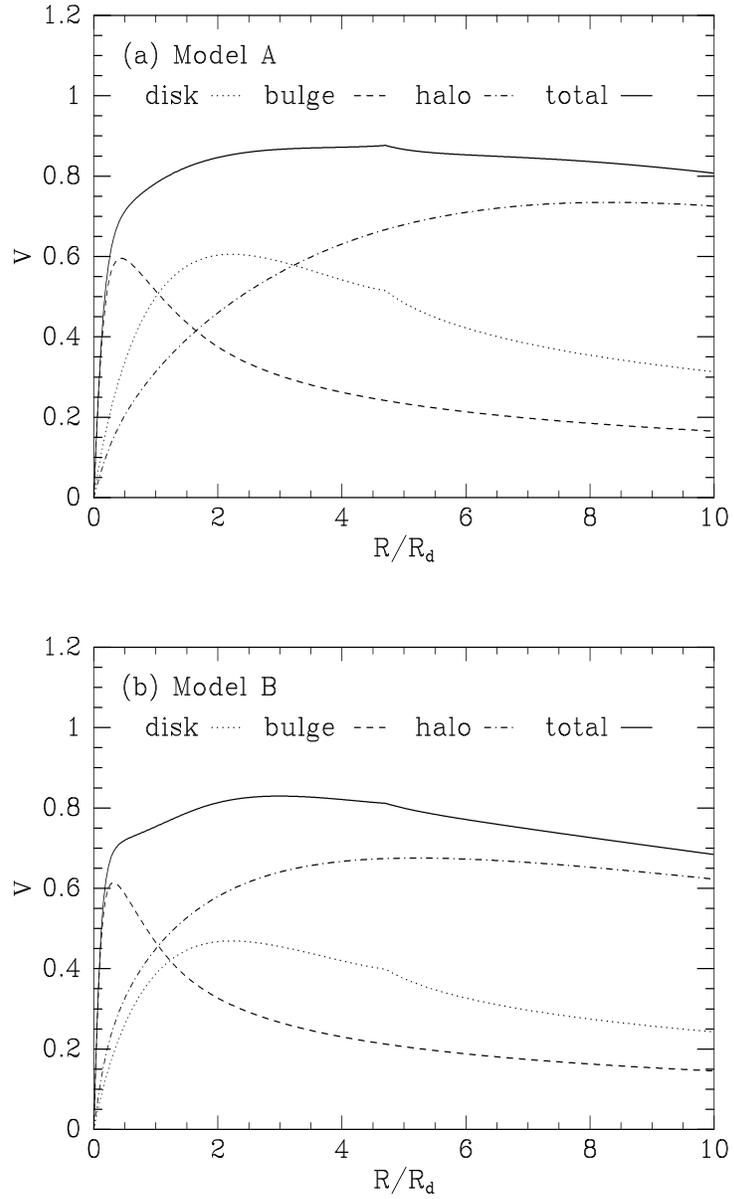

FIG. 6.–Rotation curves showing the contribution from the disk, bulge, and halo to the total for (a) Model A and (b) Model B

TABLE 1
Galaxy Model Parameters [a]

| Model | \multicolumn{4}{c}{Disk} | | | | \multicolumn{4}{c}{Bulge} | | | | \multicolumn{5}{c}{Halo} | | | | |
|---|---|---|---|---|---|---|---|---|---|---|---|---|---|
| | $\Psi_o$ | $M_d$ | $R_d$ | $z_d$ | $M_b$ | $\Psi_c$ | $\sigma_b$ | $\rho_{b,0}$ | $M_h$ | $\sigma_h$ | $q$ | $q_\rho$ | $\frac{R_c}{R_K}$ |
| | (1) | (2) | (3) | (4) | (5) | (6) | (7) | (8) | (9) | (10) | (11) | (12) | (13) |
| RIGID | -3.0 | 0.020 | 0.10 | 0.03 | – | – | – | – | 0.59 | 0.71 | 0.8 | 0.5 | 1.0 |
| A | -2.0 | 0.095 | 0.15 | 0.02 | 0.027 | -1.0 | 0.35 | 100 | 0.61 | 0.71 | 0.9 | 0.6 | 0.1 |
| B | -1.9 | 0.057 | 0.15 | 0.02 | 0.021 | -1.0 | 0.35 | 200 | 0.45 | 0.60 | 1.0 | 0.9 | 0.1 |
| C | -1.9 | 0.057 | 0.15 | 0.02 | 0.022 | -1.0 | 0.35 | 200 | 0.44 | 0.60 | 0.9 | 0.6 | 0.1 |

[a] (1) central potential, (2) disk mass, (3) disk scale radius, (4) disk scale height, (5) bulge mass, (6) bulge cut off potential, (7) bulge velocity dispersion, (8) bulge central density, (9) halo mass, (10) halo velocity dispersion, (11) halo potential flattening, (12) halo density flattening, (13) core radius over King radius. For the halo, $\rho_1 = 1$.

### 3.2 Static Bulge/Halo Potential

We first followed the disk evolution in a static bulge/halo potential with Models A and B (a fast and a slow precession) as a warm-up exercise to see if we could reproduce the predicted steady precession of the warped mode. We inserted a warp into the disk according to the SC linear mode calculations.

We measured the inclination and the longitude of the disk line of nodes as a function of time in four radial bins by calculating the principal moments of the normalized inertia tensor, $I_{ij} = \sum x_i x_j / r^2$, of the particles in each bin (Dubinski & Christodoulou 1994). Models A and B both begin to precess at the predicted rate and for the most part maintain their warp. Figure 7 presents the evolution of the Model A from an edge on view of the disk in the precessing frame. The disks in models A and B maintain the warped shape though there is a gradual decay of the inclination of the disk at all radii (Figure 8). This decay probably results from the gradual heating of the disk from both the development of internal spiral instabilities and the discreteness of the N-body simulations. Model A also develops a significant bar mode within two scale radii by the end of the simulation which may also be affecting the evolution. Model B does not develop a bar. Also, the warp mode predicted by ring models is not necessarily the exact mode of a warm, stellar disk. The ring approximation is especially poor in the center of the disk, where the velocity dispersion is a significant fraction of the circular speed. At the disk edge, the ring approximation also becomes less accurate since the density is poorly sampled and some particles are found to decouple from the disk.

Despite the slow decline in the inclination, the precession rate of the N-body disks agrees with the warp mode predictions. Figure 9 shows the evolution of the longitude of the line of nodes (the precessional phase angle) for both models. The precession rate is constant and agrees with predicted values.

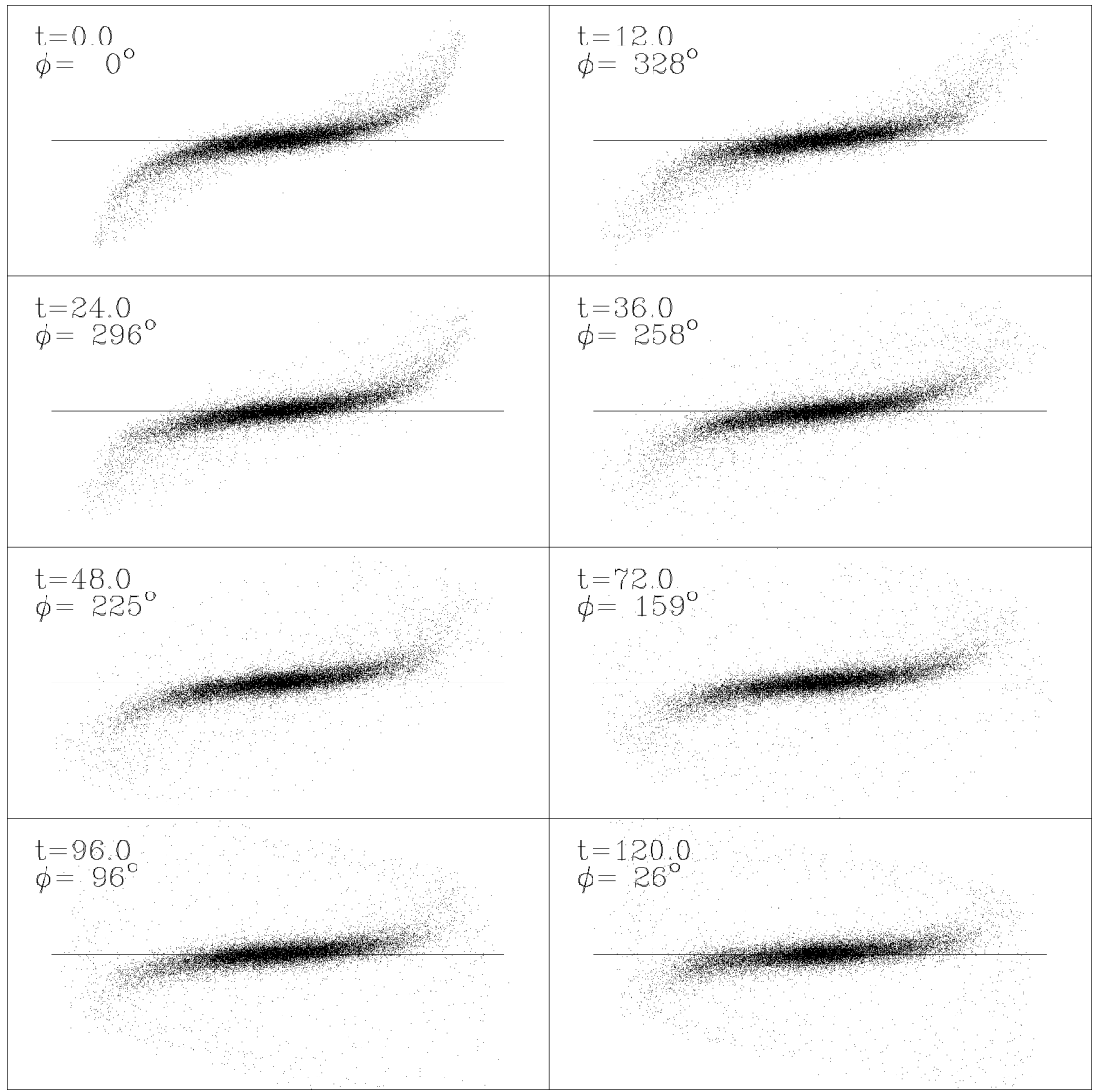

FIG. 7.–Edge-on view of the evolution of the warped mode of Model A as viewed in the precessing frame of the disk. The simulation time and longitude of the disk line of nodes, $\phi$, are shown for each snap shot.

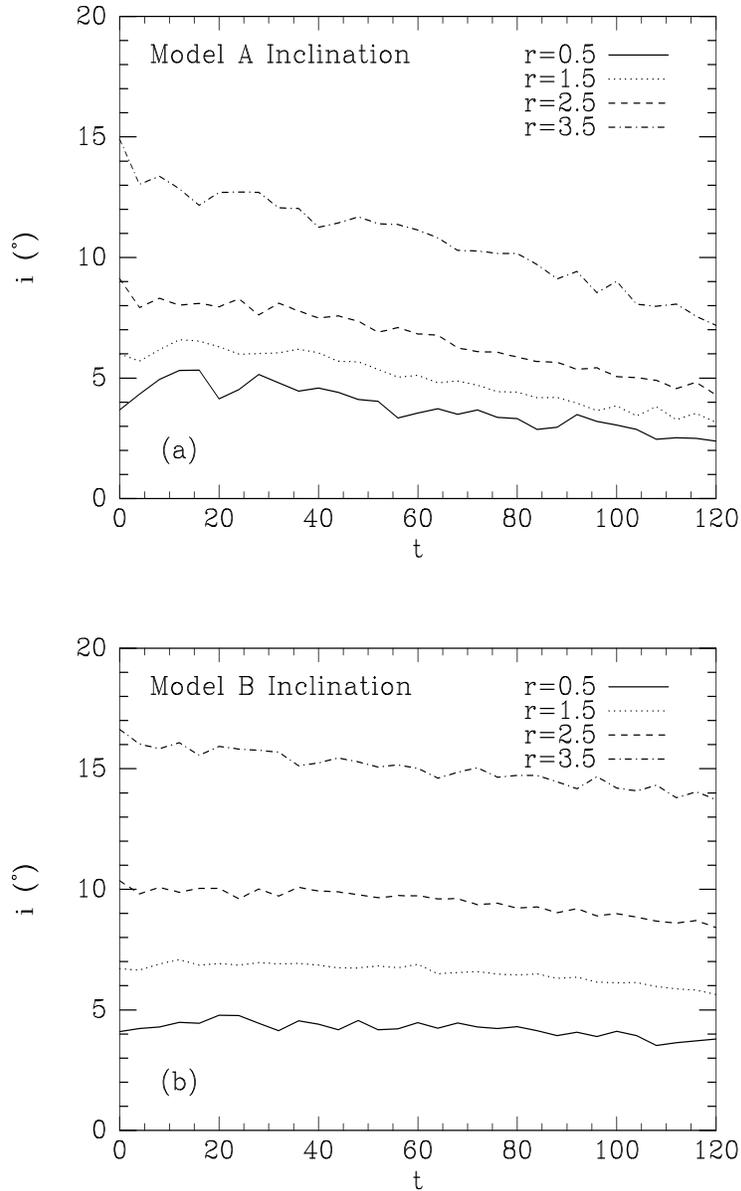

Fig. 8.—Evolution of the inclination of the warped disk as measured by the inertia tensor in 4 radial bins for (a) Model A and (b) Model B. The disk consists of $N$-body particles while the halo and bulge potentials are kept fixed. The inclination of the disk decreases steadily, probably as a result of the gradual heating of the disk by discreteness effects. Model B does not decay as quickly because the disk has a lower mass and heating is not as rapid. Nevertheless, the warped modes retain their integrity for a full precession period in Model A and a third of a precession period for Model B.

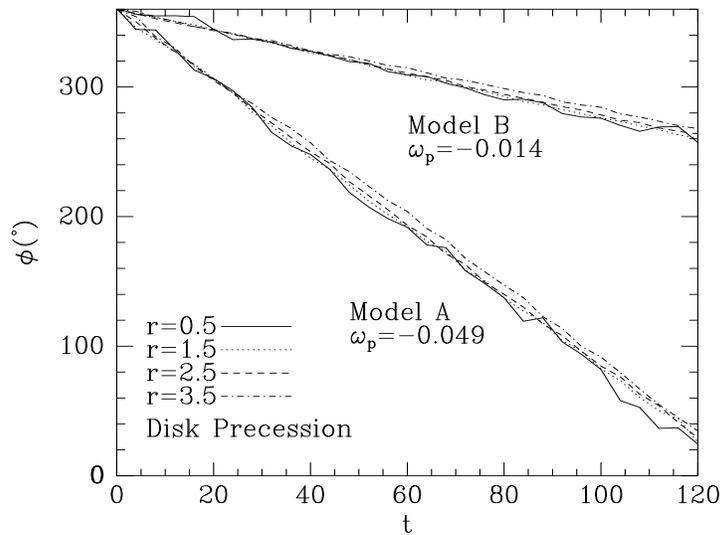

FIG. 9.—Disk precession as shown by the evolution of the longitude of the disk line of nodes, $\phi$, for (a) Model A and (b) Model B. As in figures 7 and 8, the halo and bulge potentials are static. The 4 curves for each model refer to the precessional phase, $\phi$, in each of the four radial bins seen in figure 8. The precession frequencies as measured by a least-squares fit are also shown and agree with the theoretical values.

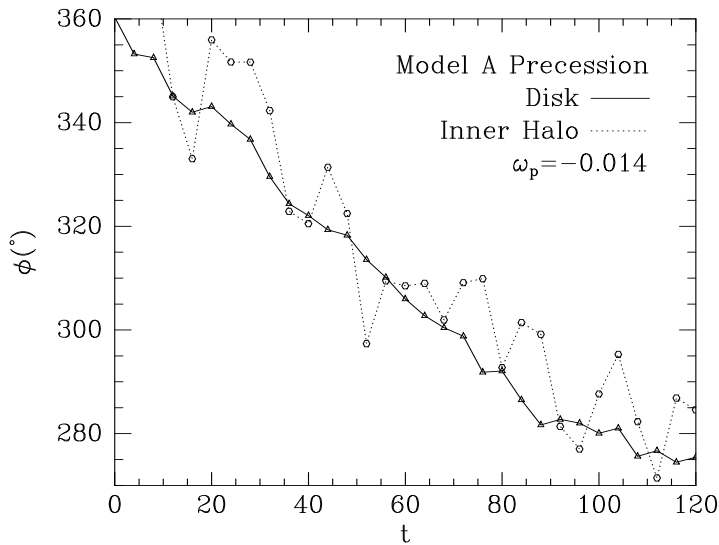

Fig. 10.—The evolution of the longitude of the disk line of nodes for Model A as measured by the $r = 1.5$ bin in comparison to the precession of the inner halo as measured by the orientation of the halo and bulge particles within 3 scale radii. The halo, bulge and disk now consist of N-body particles. The precession is considerably slower $\omega_p \approx -0.014$ then the linear-mode value of $\omega_p = -0.049$ as the bulge and halo become aligned with the disk. The torque causing the precession is exerted by the outer halo which is still misaligned at late times.

These results show that the rigid ring warp mode calculations are valid for collisionless stellar disks when we assume the halo and bulge potential are static. We now investigate fully dynamic models.

### 3.3 Dynamic Bulge and Halo

In our final simulations, we allow the halo and bulge to react to the disk's motion. We leave the initial potential of the bulge and the halo unchanged, but now we replace the background potential with particles, sampled from the distribution function. Figure 10 shows the disk precession for Model A as represented by the $r = 1.5$ radial bin along with the inner halo precession. We calculated the alignment of the halo using the method described in Dubinski & Carlberg (1991). The principal axes and axial ratios of the ellipsoidal shells are calculated by diagonalizing the normalized inertia tensor, $M_{ij} = \sum x_i x_j / a^2$, where $a$ is the elliptical radius, $a^2 = x^2 + y^2/q_1^2 + z^2/q_2^2$ and $q_1$ and $q_2$ are the axial ratios of the ellipsoidal shell. The response of the halo and bulge is immediate. Within one orbital period ($T \sim 13$ units), the precession of the disk slows down considerably in Model A. For Model B, the bulge and halo respond so quickly that no measurable precession occurs at all. The inner halo aligns rapidly with the disk in a few orbital periods. Figure 11 shows the evolution of the inclination of the minor axis of the bulge and halo particles within 3 disk scale radii in

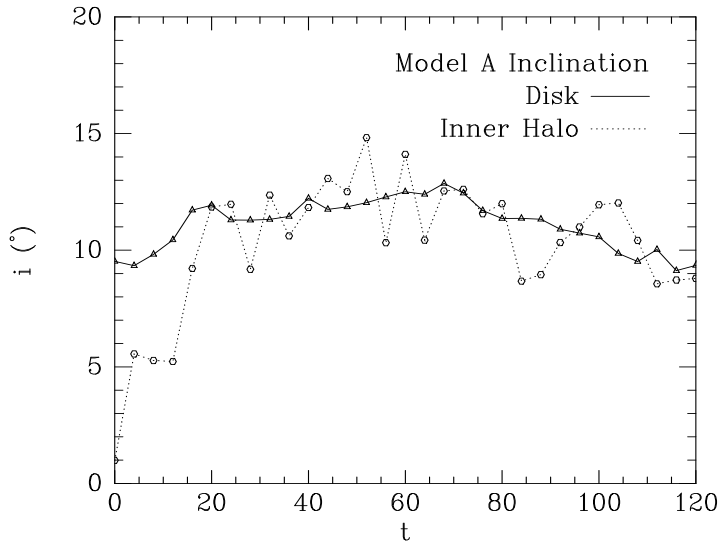

Fig. 11.—A comparison of the evolution of the disk inclination and the inclination of the inner halo and the bulge. The bulge and inner halo and disk align with each other within a few orbital periods.

Model A. The halo rapidly aligns with the disk and slows down the precession. The disk continues to precess, however, since the outer halo maintains its original orientation. The torque driving the precession is provided by this outer asymmetry. Despite the dramatic orbital readjustment in the disk there is very little change in the monopole terms of the potential. The spherically averaged density profile remains virtually unchanged during the relaxation to equilibrium (Figure 12).

Finally, we set up a Model C as a case of highly inclined system ($i = 45°$) with fast precession like Model A. Furthermore, we *do not* include the warping mode of the disk partially because of the high inclination but also to see if the disk would settle to a warped state. Under the assumption of a rigid bulge/halo potential, the disk would quickly twist and warp out of shape, attempting to settle into a (highly warped, in this case) mode shape. However, in our dynamic-halo simulation we find again that the halo responds very rapidly to the disk, slowing down the precession and therefore preventing any internal twisting (Figure 13). Figure 14 and 15 shows the evolution of the precession and inclination of the disk in comparison to the inner and outer halo halo for this highly inclined case. The inner halo was defined as the mass of the bulge and halo with $r < 3.0$ and the outer halo was defined as all mass in the ellipsoidal shell in the range $5.0 < r < 10$. Again, the inner halo and bulge align with the disk within 5 orbital periods and slow down the precession (Figure 14). The outer halo also shows a gradual evolution, as the effect of the inner alignment spreads outward. The precession essentially stops by $t \approx 120$ and reverses a few degrees though this is probably a result of measurement error. The inclination of the disk declines to $i = 20°$ by $t = 168$ at which time the halo is aligned with the disk out to ten disk scale radii. Evidently, it takes about three local orbital periods for the disk and halo to align.

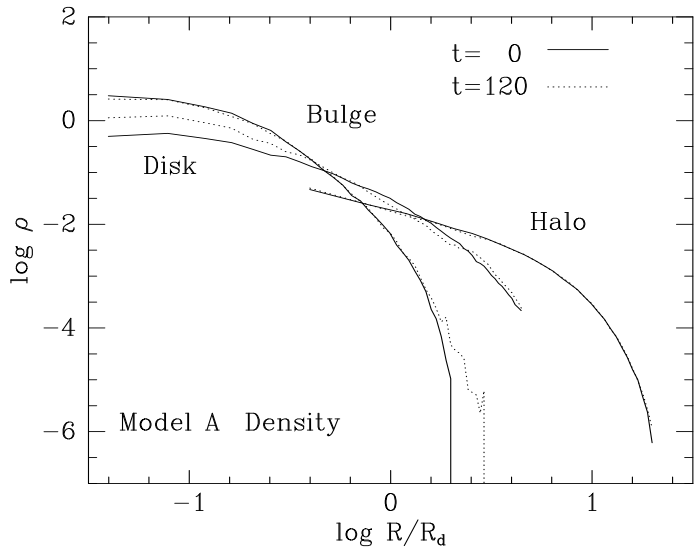

Fig. 12.–The initial and final spherically averaged density profiles for the disk, the bulge and the halo in Model A. Despite the realignment of the inner bulge and halo the spherically averaged density profiles do not change dramatically.

As the precession slows down, the shape of the disk changes. Near the end of the simulation, the outer regions bend down towards the plane of the halo, as is characteristic of a slow precession mode. This bend is not very long-lived in the simulation because by this stage the halo is almost aligned with the disk, so there is little torque affecting the disk. It is possible that this is partly an effect of our model: our halos get rounder at large radii and are truncated at about 20 disk scale radii, so once the inner portion has aligned there is no longer a reservoir of dark halo particles which might continue to provide a torque. Real halos may extend to 60 disk scale radii or more (Fich & Tremaine 1991), so it is important to examine the possible effect of the outer regions.

As an extreme model, consider a disk galaxy in an infinite, flat-rotation curve halo. We assume that at a certain time $t$, the halo within a radius $R_a$ has aligned completely with the disk, and contributes no torque to the disk. According to our simulations, $t$ is about three orbital periods at radius $R_a$. The dark matter at larger radii is taken to be undisturbed. An upper limit to the torque it can exert on the disk is obtained by assuming that the outer halo also forms a disk, misaligned with the stellar disk inside. It is then simple to show that the inner disk experiences a tidal potential from the outer halo, of size

$$\Psi_{OH} = \frac{V_c^2}{8}\left(\frac{2z^2 - R^2}{R_a^2}\right). \tag{6}$$

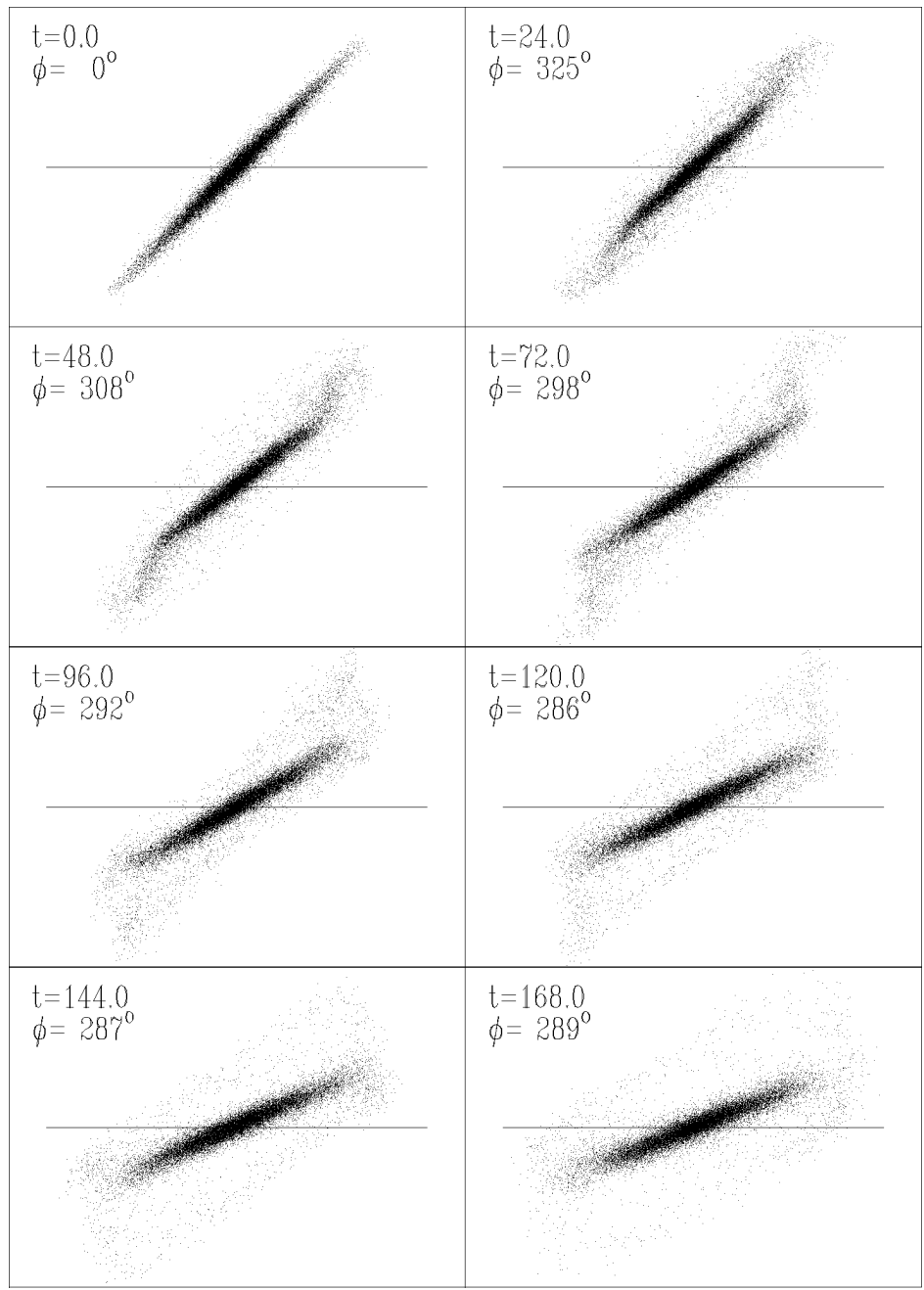

FIG. 13.—Evolution of a highly inclined disk (Model C) in a dynamic bulge and halo potential. The precession stops by $t \sim 120$. The disk holds together despite the large inclination and differential precession. Note also how at late times, after the precession has slowed, the disk takes on a slow warp shape, bending down towards the halo equatorial plane.

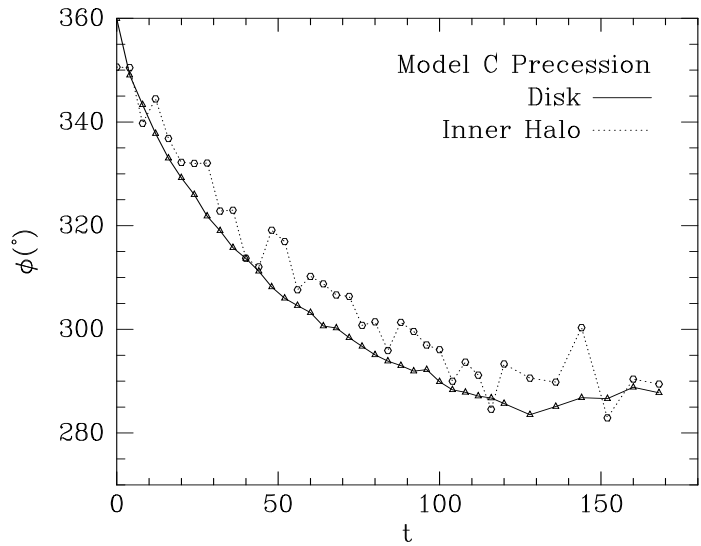

Fig. 14.—The evolution of the longitude of the disk and inner halo ($R < 3$) line of nodes for Model C. The rate of precession gradually declines as the disk, bulge and halo align with one another.

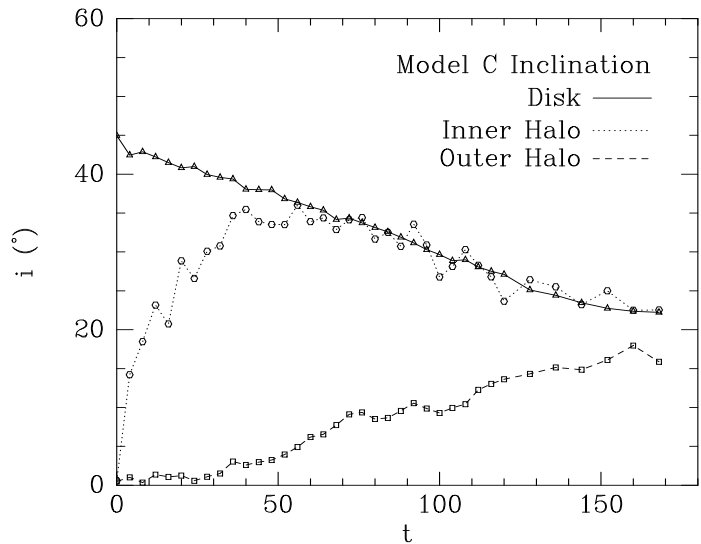

Fig. 15.—The evolution of the inclination of the disk, and the inner ($R < 3$) and outer ($5 < R < 10$) halo. The inner halo aligns with the disk within 5 orbital periods, while the outer halo takes much longer and has not quite aligned by the end of the simulation.

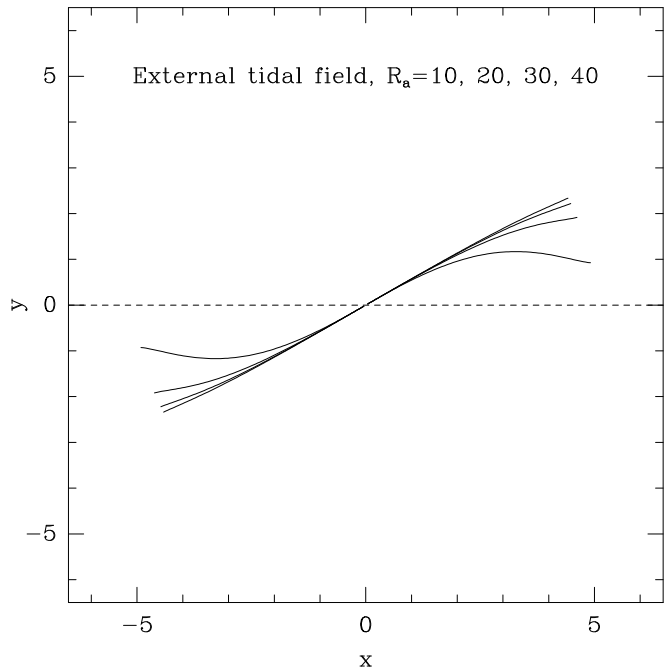

FIG. 16.–The warp-mode shapes of an exponential disk, misaligned by 30° in a uniform tidal field, as may be contributed by a distant flattened halo. The case of most extreme bending is shown, with the halo beyond radius $R_a$ taken to be a misaligned disk. The disks are truncated at 5 scale radii, and have unit mass. The halo circular speed is unity. The precession frequencies, in units of the orbital frequency at one scale radius, are $-0.018$, $-0.0048$, $-0.0022$, and $-0.0012$ for $R_a=10$, 20, 30 and 40. The warps are strongest when $R_a$ is small. Real disks may be expected to have aligned with their halos out to about 20 scale radii at the present epoch.

In an external tidal field, a disk always has a bending mode, (Toomre 1983). Shapes for an exponential disk truncated at five scale radii in such a tidal field are shown in Figure 16, for different settling radii $R_a$. As the halo and disk align out to larger and larger radii, the warp mode shape rapidly flattens.

If observed warps are indeed a response to the outermost halo before it too aligns itself with the disk, then our simulations and Figure 16 show that such warps will rapidly die out, indeed most of them should have disappeared already (if scaled to our Galaxy, where the orbital time is $\sim 0.2$ Gyr, the time for the halo to align with the disk out to 20 scale radii ($\sim 70$ kpc) is only $\sim 6$ Gyr). It appears that a mechanism needs to be found which can continually excite, or at least slow down the decay of, disk warps.

We are forced to conclude that warps envisaged as modes in rigid potentials are not likely to exist in real galaxies. A disk inclined with respect to a flattened dark halo raises strong wakes in the halo, as a result of which such a system will relax to the equilibrium state of alignment. For realistic ratios of mass for the disk, the bulge and the halo, as represented by our models, the relaxation to equilibrium occurs within a few orbital times of the disk.

Halos more extended than those we have considered barely help save the warp.

## 4 Discussion

Our results show how the inner halo rapidly realigns with the disk, eliminating the mutual torque and therefore halting the disk precession. A disk inclined with respect to a halo raises strong wakes. The natural relaxation timescale to equilibrium is just the orbital timescale in the inner regions, where the mass of the disk and halo components are comparable.

Our simulations show that in the inner regions of galaxies, the disk, the bulge and halo will share their symmetry plane. However, the outer halo may still be misaligned, leaving an external quadrupole on the disk which may still cause it to precess. Toomre's (1983) original suggestion of a torque exerted by an external ring of satellites may be an accurate description of the truth, though perhaps the external ring should be considered instead as a set of misaligned ellipsoidal mass shells at large radii. Even in this case, the warp and the precession will be subject to decay through dynamical friction and disk-halo alignment from the inside out.

We should emphasize that our simulations do not start from an equilibrium state. While the initial models (with the disk in the equatorial plane) are in equilibrium, our subsequent tilting of the disk disturbs the system. Since this change does not affect the monopole of the potential, it is not surprising to discover that the spherically averaged density profile of the system as a whole does not change as the system evolves (Figure 12). Only the azimuthal components of the orbits are modified by tilting the disk. A more realistic model might consist of a disk which grows slowly inside a pre-existing halo. However, in this case the evolution will be adiabatic, and the final state will be a disk and halo with a common symmetry plane (Binney & May 1986). Only fairly violent initial conditions, in which the disk and halo form hierarchically from sub-clumps and exchange angular momentum up to the merger of the last few pieces, will generically produce misaligned disk-halo systems (Katz & Gunn 1991). There is then no guarantee that the resulting system will be in equilibrium, and so our simulations may be appropriate in such a picture.

Some caveats on the applicability of these models to real galaxies should be mentioned. The bulges and halos described here have no rotation or radially anisotropic velocity distributions, unlike the expectation for halos formed in a cosmological collapse (van Albada 1982; Dubinski 1992). The coupling of the disk to a different distribution of orbits may lead to a slightly different settling time. In the most plausible case, where the halo rotates in the same direction as the disk, dynamical friction might be expected to be stronger than in the non-rotating case (since the precession is retrograde), leading to faster disk settling. Radial anisotropy will also affect the timescale through the number of orbits in resonance with the disk precession.

In view of the rapid alignment of the inner disk, bulge and halo, precessing warped modes within flattened halos may not exist at all. Just as isolated (halo-less) disks do not admit warp modes except for unrealistic boundary conditions (Hunter & Toomre 1969,

Sparke & Casertano 1988), it is possible that when a disk-bulge-halo system is viewed in its entirety it does not have a neutral bending mode either. There certainly is no argument for the modes of such a system to be neutral, as there was in the case of a thin disk in a static halo. Only a full-blown mode analysis of a fully responsive system can definitely settle the issue. In this context, it is interesting that Weinberg (1994) has found some very weakly damped dipole modes of spherical King (1966) models.

Alternative theories propose late cosmological infall to the exterior of galaxies as the cause of the warps. Late cosmological infall onto a galaxy does not necessarily have the same angular momentum orientation as the disk (Ryden 1988; Quinn & Binney 1992) —instead the angular momentum of material accreted at time $t$ decorrelates on a timescale $O(t)$. Most of the accreted material at late times will be dark matter, which then has the effect of slowly changing the halo angular momentum vector. As the halo slews around, the disk responds and in so doing warps. While appealing, this theory has been tested comparatively little against observations. A cosmological origin of warps does perhaps promise new insights into the galaxy formation process.

## 5  Conclusions

Our simulations emphasize the strong gravitational coupling between the disk, the bulge and the halo in the inner galaxy. If the disk is tilted with respect to the halo, the two components will align within a few dynamical times, even for unrealistically low-mass disks. For realistic disks, the disk:halo mass ratio is comparable within the radius of the optical disk so the governing timescale for the disk and inner halo to align is just the orbital time of the inner galaxy. These results answer the concerns of warp theorists about the importance of dynamical friction for precessing disks in flattened halos: the answer is that dynamical friction is very important and cannot be neglected! For this reason, it seems unlikely that galactic warps are neutral bending modes of a disk tilted inside the static potential of a flattened dark halo. Inferred values of the core radius and flattening of dark halos from observed warps should be viewed with caution. The observed correlation that galaxies with small rotation curve core radii are usually flat, whereas galaxies with large core radii often are warped (Bosma 1991), argues in favor of significant dynamical friction: the settling timescale is governed by the amount of halo matter there is in the central regions of the disk, with dense halos slowing the disk down more efficiently. However, the warp mode picture also is consistent with this correlation, since halos with small core radii call for fast precession, and fast warp modes, if they exist at all, disappear at quite small amplitudes (Kuijken 1991).

The simulations also show, not surprisingly, that the outer halo takes much longer to align with the disk. Before it settles completely, it may still exert a strong enough external torque to force the disk to precess and warp. However, if this is to be viewed as a normal bending mode, the appropriate "rigid" halo model only contributes at large radii. The inner region is aligned with, and precesses with, the disk, and does not contribute any torque. Even optimistic models for the torque that the disk may feel from the outer halo after 6 Gyr of disk-halo alignment fall somewhat short of what is required to explain the ubiquity of galactic warps.

A modified explanation for galactic warps as bending modes might be as follows. A galactic disk forms within its dark halo at some random inclination. The disk begins to precess and the inner disk, bulge and halo rapidly align shortly thereafter. As the precession slows, the coupling to outer halo orbits takes hold and causes the outer halo to align. The alignment of the halo moves out like a wave until the precession stops completely. The observed disk warps may simply be the quasistatic response as the outer halo aligns itself with the disk, not unlike the gentle slewing of the disk expected for gradual secondary infall by Ostriker & Binney (1989). Warped disks may simply be the dying echoes of an asymmetric beginning. The timescale for complete realignment nevertheless appears to be a little shorter than a Hubble time, in which case a mechanism is called for which excites fresh bending waves and/or slewing.

One strong prediction of this effect is that the isophotes measured in the X-ray halos around disk galaxies (warped or not) should be aligned in the center and only misaligned at the extremes.

### Acknowledgments

JD acknowledges the support of a CfA Postdoctoral Fellowship and KK acknowledges the support of a Hubble Fellowship through grant HF-1020.01-91A awarded by the Space Telescope Science Institute (which is operated by the Association of Universities for Research in Astronomy, Inc., for NASA under contract NAS5-26555).

### Appendix: Disk, Bulge and Halo Initial Conditions

#### A.1 Halo

The halo initial conditions are derived from the distribution function of the lowered Evans model (Evans 1993; Kuijken & Dubinski 1994). This function is the flattened analogue of a King model (King 1966). It depends on the energy, $E$, and $z$-component of angular momentum, $L_z$ and is given by,

$$f(E, L_z) = \begin{cases} [(AL_z^2 + B)\exp(-E/\sigma_h^2) + C][\exp(-E/\sigma_h^2) - 1] & \text{if } E < 0, \\ 0 & \text{otherwise.} \end{cases} \qquad (7)$$

where,

$$A = \frac{8(1-q^2)G\rho_1^2}{\pi^{1/2}q^2\sigma_h^7}, \qquad B = \frac{4R_c^2 G\rho_1^2}{\pi^{1/2}q^2\sigma_h^5}, \qquad \text{and} \qquad C = \frac{(2q^2-1)\rho_1}{(2\pi)^{3/2}q^2\sigma_h^3}. \qquad (8)$$

The free parameters are the halo flattening, $q$, a core radius, $R_c$, a density factor, $\rho_1$, and a velocity dispersion, $\sigma_h$, as well as the central value of the potential, $\Psi_o < 0$. One can derive the density as a function of $R$ and the potential $\Psi$ giving,

$$\rho_h(\Psi, R) = \frac{1}{2}\pi^{3/2}\sigma_h^3(AR^2\sigma_h^2 + 2B)\operatorname{erf}(\sqrt{-2\Psi}/\sigma_h)\exp(-2\Psi/\sigma_h^2)$$
$$+ (2\pi)^{3/2}\sigma_h^3(C - B - AR^2\sigma_h^2)\operatorname{erf}(\sqrt{-\Psi}/\sigma_h)\exp(-\Psi/\sigma_h^2)$$
$$+ \pi\sqrt{-2\Psi}[\sigma_h^2(3A\sigma_h^2 R^2 + 2B - 4C) + \frac{4}{3}\Psi(2C - A\sigma_h^2 R^2)], \quad (9)$$

where $\operatorname{erf}(x) = 2\pi^{-1/2}\int_0^x \exp(-t^2)\,dt$ is the usual error function. In Table 1, we let $\rho_1 = 1$ for all the models, and we quote $v_o = 2^{1/2}\sigma_h$.

### A.2 Bulge

For the bulge we use a King (1966) model modified slightly to have an energy cut-off at $E = \Psi_c$.

$$f(E) = \begin{cases} \rho_1(2\pi\sigma_b^2)^{-3/2}\exp[(\Psi_o - \Psi_c)/\sigma_b^2]\{\exp[-(E - \Psi_c)/\sigma_b^2] - 1\} & \text{if } E < \Psi_c, \\ 0 & \text{otherwise.} \end{cases} \quad (10)$$

where $\rho_1$ is a density scale factor, and $\sigma_b$ is the bulge velocity dispersion. The bulge density is only a function of $\Psi$ given by,

$$\rho_b(\Psi) = \rho_1 \exp\left(\frac{\Psi_o - \Psi_c}{\sigma_b^2}\right)\left\{\exp\left(\frac{\Psi_c - \Psi}{\sigma_b^2}\right)\operatorname{erf}\left(\sqrt{\frac{\Psi_c - \Psi}{\sigma_b^2}}\right) - \sqrt{\frac{4(\Psi_c - \Psi)}{\pi\sigma_b^2}}\left(1 - \frac{2}{3}\frac{(\Psi_c - \Psi)}{\sigma_b^2}\right)\right\}.$$
$$(11)$$

### A.3 Disk

For the disk we do not have a distribution function so we use the explicit function for the density after Hernquist (1993),

$$\rho_d(R, z) = \frac{M_d}{4\pi R_d^2 z_d}\exp(-R/R_d)\operatorname{sech}(z/z_d)^2. \quad (12)$$

where $M_d$ is the disk mass, $R_d$ is the disk scale radius, and $z_d$ is the disk scale height. The disk is also cut off abruptly at $R = 4.7 R_d$. We use the recipe of Hernquist (1993) derived from the collisionless Boltzmann equation to set up the disk velocities. The radial, tangential and $z$ velocity dispersions are approximated by,

$$\sigma_R^2 \propto \exp(-R/R_d), \qquad \sigma_\phi^2 = \sigma_R^2 \frac{\kappa^2}{4\Omega^2}, \qquad \text{and} \qquad \sigma_z^2 = \pi G \Sigma(R) z_d \quad (13)$$

where $\kappa$ is the radial epicyclic frequency, $\Omega$ is the orbital frequency, both functions of $R$, and $\Sigma(R)$ is the surface density of the disk. The constant of proportionality for $\sigma_R^2$ is chosen so that the disk satisfies the Toomre (1963) stability criterion, $\sigma_R > \sigma_{R,crit} \equiv 3.36 G\Sigma/\kappa$. For the disk simulations, we choose a Toomre $Q$ parameter equal to 1.5 i.e. $\sigma_R = 1.5\sigma_{R,crit}$. Finally, we assign the mean tangential velocity using,

$$\bar{v}_\phi^2 - v_c^2 = \sigma_R^2\left(1 - \frac{\kappa^2}{4\Omega^2} - 2\frac{R}{R_d}\right) \quad (14)$$

We select velocities assuming they are locally Gaussian variates of the given dispersions.

*A.4 Calculating the Total Potential*

The last stage is finding the solution to Poisson's equation for the net potential:

$$\nabla^2 \Psi = 4\pi G[\rho_d(R,z) + \rho_b(\Psi) + \rho_h(R,\Psi)]. \tag{15}$$

The starting conditions are $\Psi = \Psi_o$, with $\partial\Psi/\partial r$ and $\partial\Psi/\partial z$ zero at the origin. We use the iterative multipole method described in Kuijken & Dubinski (1994) and solve the equations to order $l = 16$. Once we have the potential, we realize each component in particles and use them for simulations.